\documentclass[aps,pra,reprint,superscriptaddress]{revtex4-2}

\usepackage{graphicx}
\usepackage{amsmath,amssymb,amsfonts}
\usepackage{hyperref}
\hypersetup{colorlinks = true, urlcolor=blue, linkcolor = blue, citecolor=blue}

\DeclareMathOperator{\Tr}{Tr}

\begin{document}

\title{Populating topologically protected edge states of a Chern insulator\\ with the cold-atom elevator scheme and measurements}

\author{Toke Marstrand Pontoppidan Lindhard}
\affiliation{Department of Physics and Astronomy, Aarhus University, DK-8000 Aarhus C, Denmark}
\affiliation{ POLIMA---Center for Polariton-driven Light--Matter Interactions, University of Southern Denmark, DK-5230 Odense M, Denmark}

\author{Anne E. B. Nielsen}
\affiliation{Department of Physics and Astronomy, Aarhus University, DK-8000 Aarhus C, Denmark}

\begin{abstract}
Two-dimensional Chern insulators support topologically protected, chiral edge currents, and these can be detected in experiments with ultracold atoms in optical lattices. It has previously been shown that one can populate a selected group of edge states of a Chern insulator by transferring particles from a reservoir. Here, we numerically investigate the effect of performing an instantaneous, projective measurement on the reservoir before the reservoir is discarded. In this way, the final state of the system is pure and described by a wavefunction. We also show that quite likely measurement outcomes can help to increase the final number or percentage of particles in the chiral edge states through postselection. Without the measurement step, the physics can be described in terms of single-particle physics. The measurement significantly complicates the description. By appropriately rewriting the analytical expressions, we show that measurement probabilities, expectation values, averages of expectation values, and purity can nevertheless be computed from the state before the measurement in a way that scales only linearly with the number of lattice sites for a fixed number of particles. This enables us to investigate a setup with, for instance, 14 particles and 198 lattice sites numerically. The approach applies generally to noninteracting, fermionic models that conserve the number of particles. 
\end{abstract}

\maketitle

\section{\label{sec:introduction}Introduction}

Two-dimensional Chern insulators are important examples of topology in quantum systems \cite{PhysRevLett.61.2015,RevModPhys.82.3045,BERGHOLTZ_2013}. They are band insulators with broken time reversal symmetry and non-zero Chern number. Their edges host topologically protected edge states that do not allow backscattering from local perturbations as long as the bulk gap does not close \cite{PhysRevLett.49.405,PhysRevLett.51.51,PhysRevB.31.3372}. They have inspired several applications such as topological lasers \cite{StJean2017,Bandres2018} and topological beam splitters \cite{WangSu2017,Qi2021} as well as applications within 5G communication devices \cite{Kumar2022,Nagulu2022}.

Ultracold atoms in optical lattices provide a versatile platform for quantum simulations \cite{RevModPhys.80.885, Lewenstein_2007}, and they can also simulate Chern insulators \cite{PhysRevLett.111.185301,miyake2013realizing}. This is achieved by creating an artificial gauge field, mimicking a magnetic field \cite{RevModPhys.83.1523}. The Chern number has been measured to detect the topology experimentally \cite{jotzu2014experimental, aidelsburger2015measuring}. In these early experiments, the systems did not have a sharp edge, and this prevented the detection of topological edge states. The dynamics of particles occupying the topological edge states were observed only recently \cite{Braun_2024,Yao_2024,hesse2025probing}. This was made possible by the development of optical box potentials that allow for the engineering of sharp edges \cite{Navon_2021,PhysRevLett.110.200406,Chomaz2015, PhysRevLett.118.123401, PhysRevLett.120.060402, PhysRevResearch.3.033013}. In these systems, decay to lower-lying states is slow, and therefore chiral dynamics can be seen even without populating the states below the edge states. In a recent theoretical paper \cite{cold_atom_elevator}, it was proposed how one can use a cold-atom elevator to inject particles from a reservoir into the edge states of a Chern insulator at specific energies. In this scheme, it is not known exactly how many particles are being transferred and the final state of the system alone is a highly mixed state described by a density operator. Considering the system alone, quantum coherence is thus lacking.

Here, we numerically investigate the preparation of pure system states that primarily occupy the edges and have a fixed total number of particles. As in \cite{cold_atom_elevator}, we consider injection of particles from a reservoir, but instead of discarding the reservoir after the injection, we perform a projective measurement of the reservoir, which disentangles the reservoir and the system and leads to a pure system state described by a wavefunction.  Specifically, we start from a topologically trivial reservoir, which is prepared in the many-body ground state with a fixed and known number of particles. The system is initially empty. The reservoir and the system are then coupled along the edge. After a certain time, the coupling is turned off, but the state of the two parts is now entangled. We disentangle the system and the reservoir by simulating a measurement of the number of particles on each site of the reservoir and computing the state of the system given the outcome of the measurement. Such measurements can be achieved with a quantum gas microscope \cite{bakr2009quantum,gross2021quantum}. For simplicity, we assume that the measurement step occurs instantly, although the measurements do take time in experiments. After the measurement, the system is in a pure state, and the number of particles in the system is fixed to the initial number of particles minus the number of particles detected in the reservoir. For suitably chosen parameters, the majority of these particles occupy the edge states. Our numerical simulations show that for some, quite likely, measurement outcomes, the number of particles in the edge states is increased compared to the case without a measurement. Postselection can hence also be used as a tool to enhance the transfer.

Without the measurement step, the many-body physics can be formulated in terms of single-particle physics, and therefore rather large system sizes can be investigated numerically. The complexity increases considerably when the measurement step is introduced. We show how one can nevertheless rewrite the expressions for measurement probabilities, expectation values, averages of expectation values, and purity in a way that allows for efficient computations. In this way, we obtain numerical results for systems with, e.g., $14$ particles and $198$ sites. The reformulations are not restricted to the Harper-Hofstadter model \cite{Harper_Hofstadter_model} considered here, but apply generally to noninteracting, fermionic models that conserve the number of particles.

The paper is organized as follows. In Sec.\ \ref{sec:Scheme}, we describe the considered model and the proposed scheme to populate the topological edge states. In Secs.\ \ref{sec:measurement} and \ref{sec:probability}, we derive the formulas that allow us to do simulations for large system sizes despite the measurement step. In Sec.\ \ref{sec:Results_and_discussion}, we present the results for the probabilities of the different measurement outcomes and the final population of the states of the system after the measurement. Finally, Sec.\ \ref{sec:Conclusion} concludes the paper.

\section{\label{sec:Scheme}Scheme for edge-state injection}

\subsection{Model}

We consider spinless fermions in the Harper-Hofstadter model \cite{Harper_Hofstadter_model} on a two-dimensional square lattice. The lattice is divided into two regions that are called system $S$ and reservoir $R$, respectively. The system and the reservoir are described by the Hamiltonian
\begin{multline}
\hat{H} = - J \sum_{\langle ij\rangle \notin \text{border}} \left( e^{i\phi_{ij}} \hat{a}_i^\dag \hat{a}_{j} + \text{h.c.} \right)\\
- J_{\text{B}} \sum_{\langle ij\rangle \in \text{border}} \left(e^{i\phi_{ij}} \hat{a}_i^\dag \hat{a}_{j} + \text{h.c.} \right) 
+ \varepsilon_{\text{R}} \sum_{i\in R} \hat{a}^\dag_i \hat{a}_i.
\end{multline}
The operator $\hat{a}_i$ annihilates a particle on site $i$ and h.c.\ means Hermitian conjugate. The first two terms of the Hamiltonian describe complex hopping between nearest neighbor sites with Peierls phases corresponding to a uniform magnetic field perpendicular to the plane of the lattice. Specifically, $\phi_{ij}$ is zero for hopping along $y$ and $\phi y/a$ for hopping along $x$, where $a$ is the lattice constant, $\phi = 2\pi \alpha$ is the magnetic flux per plaquette, and $\alpha \in \mathbb{Q}$ is a rational number. The strength of the hopping terms is $J$ for hops within the system or within the reservoir and $J_{\text{B}}$ for hops between the system and the reservoir. The third term in the Hamiltonian is a potential of strength $\varepsilon_{\text{R}}$ that allows one to shift the energies of the particles in the reservoir relative to the energies of the particles in the system. This shifting is called the cold-atom elevator. Experimentally, one can decouple the system and the reservoir by taking $|\varepsilon_{\text{R}}|$ to be sufficiently large. In the computations below, we instead decouple the two by taking $J_{\text{B}}$ to be zero, as this is computationally simpler. When the two are decoupled, one can compute energies and eigenstates separately for the system and the reservoir, and we denote these energies by $E_{S,j}$ and $E_{R,j}$, respectively. Throughout, we choose the boundary conditions to be open, as this is the most relevant case for experiments.

\begin{figure}
\includegraphics[width=\linewidth]{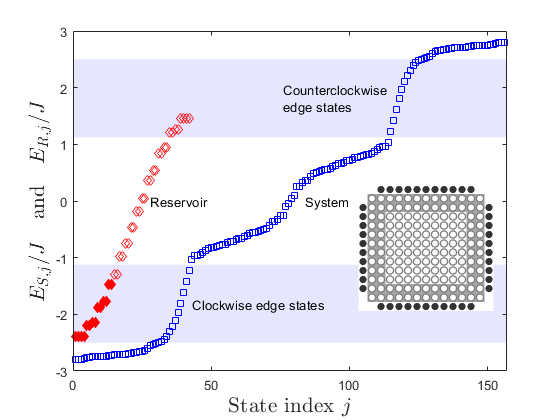}
\caption{Single-particle energy spectrum (blue squares) of the system with $13\times12$ sites (white sites in inset) and magnetic flux $\phi=\pi/2$. States with at least $97 \%$ of their particles on the two outermost rows or columns (the region with gray background in the inset) are marked as edge states. The clockwise group consists of the states $C=\{31,\ldots,42\}$, and the counterclockwise group consists of the states $CC=\{115,\ldots,126\}$. The single-particle spectrum (red diamonds) of the reservoir (black sites in the inset) with $\epsilon_R$ given in Eq.\ \eqref{er} is also shown. The filled (empty) diamonds indicate single-particle states that are initially occupied (unoccupied).}\label{fig:spectrum}
\end{figure}

Considering the single-particle spectrum of the system, one observes groups of chiral edge states in between the bulk states. The density of states is lower in regions with edge states than in regions with bulk states. The number of edge groups is determined by the strength of the magnetic flux $\phi$, and thus by $\alpha$. With the choice of $\phi=\pi/2$ (corresponding to $\alpha=1/4$), the edge states are collected in two groups with opposite chirality as seen in Fig.\ \ref{fig:spectrum}. In the following, we denote the set of clockwise edge states by $C$ and the set of counterclockwise states by $CC$. The aim is ideally to populate the clockwise edge states without populating other states in the system. 

The states of the reservoir are two-fold degenerate in Fig.\ \ref{fig:spectrum}. This happens because the reservoir consists of two copies of a $1\times 10$ system and two copies of an $11\times 1$ system. The copies are uncoupled and equivalent up to a gauge transformation. Note also that the reservoir is by itself topologically trivial, as one cannot go around a closed loop inside the reservoir.

\subsection{\label{sec:new_scheme}Scheme}

\begin{figure}
\includegraphics[width=\linewidth]{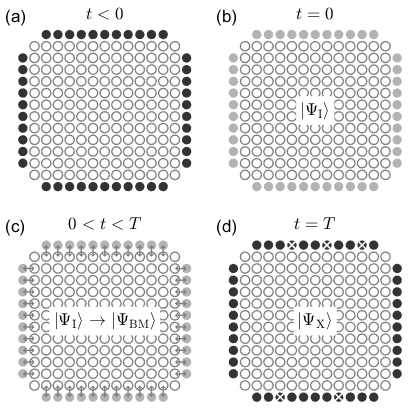}
\caption{The proposed scheme. Each circle represents a site on the two-dimensional lattice. The circles with white filling constitute the $13 \times 12$ system, while the circles with gray or black filling constitute the reservoir of sizes $1 \times 10$ and $11 \times 1$. The reservoir does not wrap around the corners of the system to avoid a too strong coupling between the corner sites and the reservoir. (a) For $t<0$, the system and the reservoir are decoupled, either by putting $J_{\text{B}}$ equal to zero or by choosing $|\varepsilon_{\text{R}}|$ to be sufficiently large. This is represented by the dark color of the reservoir sites. (b) At $t=0$, the coupling between the system and the reservoir is turned on by putting $J_{\text{B}}=J$ and instantly lifting $\varepsilon_{\text{R}}$ to the value in Eq.\ \eqref{er}. This is shown by the gray color of the reservoir sites. (c) For $0<t<T$, particles are transferred into the system through the border between the system and the reservoir as illustrated with the arrows. (d) Finally, at $t=T$, the reservoir is decoupled from the system again, and it is measured which sites in the reservoir are occupied. The occupied sites are marked by crosses. The example shown here is one of the most likely measurement outcomes among those with five particles in the reservoir.} \label{fig:scheme}
\end{figure}

The scheme is illustrated in Fig.\ \ref{fig:scheme}. The steps (a)-(c) are similar to the protocol considered in \cite{cold_atom_elevator}, although the geometry is different. At time $t<0$, the system and the reservoir are decoupled, either by putting $J_{\text{B}}=0$ or by taking $|\varepsilon_{\text{R}}|$ to be large. We assume that the system is initially empty and that the reservoir is initially in the many-body ground state with a fixed and known number of particles. 

At time $t=0$, the coupling is turned on by putting $J_{\text{B}}=J$ and choosing $\epsilon_R$ appropriately. As our aim is to populate the group of clockwise edge states, we want the range of energies of the occupied states in the reservoir to roughly fit the range of energies of the clockwise edge states. With the lowest $N=14$ single-particle states of the reservoir filled, the energy range of the occupied states is sufficiently narrow, as seen in Fig.\ \ref{fig:spectrum}. We adjust the average energy of the occupied states to the average energy of the clockwise edge states by choosing
\begin{equation}\label{er}
\epsilon_R = \frac{1}{|C|}\sum_{j\in C} E_{S,j}-\frac{1}{N}\sum_{j=1}^{N} E^{(\epsilon_R=0)}_{R,j},
\end{equation}
where $|C|$ is the number of elements in the set $C$ and the superscript $(\epsilon_R=0)$ means that $E_{R,j}$ is here computed for $\epsilon_R=0$. With the coupling turned on, particles can hop from the reservoir to the system, and as sites in the system become populated, particles can also hop in the opposite direction. The reservoir surrounds the system such that all four sides of the system couple to the reservoir. The reservoir is one site wide, which prevents particles in the reservoir to be far from the system. The reservoir does not wrap around the corners of the system to prevent a too strong coupling between the corners and the reservoir. 

At time $T$, the coupling is turned off again. The system and the reservoir are now entangled. We want to prepare a state of the system, which has a well-defined number of particles and is described by a wavefunction rather than a density operator. We achieve this with a projective measurement, which measures the number of particles on each site in the reservoir. For simplicity, we will here assume that the measurement can be performed instantly. This transforms the quantum state into a direct product of a state involving only the system sites and a trivial state of the reservoir. As the system and reservoir are no longer coupled, we can ignore the reservoir for all later times. In addition, since we know both the total number of particles and the number of particles in the reservoir, the number of particles in the system is also fixed. 

\subsection{\label{sec:time_evo}Time evolution of the state}

We next formulate the equations needed to compute the final state of the system. The initial state at $t=0$ takes the form
\begin{equation}
	|\Psi_{\text{I}} \rangle = \prod_{j=1}^{N} \hat{\zeta}_j^\dag |0\rangle = \prod_{j=1}^N \left(\sum_{m \in R} \langle 0|\hat{a}_m \hat{\zeta}_j^\dag |0\rangle\, \hat{a}^\dag_m \right) |0 \rangle,
	\label{eq:initial_state}
\end{equation}
where $\hat{\zeta}_j^\dag$ creates a fermion in the $j$th single-particle energy eigenstate of the reservoir and the eigenstates are numbered from low to high energy. Time evolving the state from $t=0$ to $t=T$, we get the state right before the measurement
\begin{equation}
	|\Psi_{\text{BM}} \rangle = e^{-i\hat{H} T} |\Psi_{\text{I}} \rangle.
	\label{eq:time_evolution_operator}
\end{equation}
This can also be expressed as
\begin{equation}
|\Psi_{\text{BM}} \rangle = \prod_{j=1}^{N} \hat{\psi}_j^\dag |0\rangle = \prod_{j=1}^N \left(\sum_{m \in S \cup R} \langle 0|\hat{a}_m \hat{\psi}_j^\dag|0\rangle\, \hat{a}^\dag_m \right) |0 \rangle
\label{eq:mult_particle_before_measurement}
\end{equation}
with
\begin{equation}
\hat{\psi}_j^\dag=e^{-i\hat{H}T}\hat{\zeta}_j^\dag e^{i\hat{H}T}.
\end{equation}
Note that the states now extend over both the system and the reservoir. 

The measurement is described in terms of a projection operator that acts on the sites of the reservoir. Specifically, the projection operator for measuring one particle on each of the sites $X = \{ x_1, \dots, x_n \}$ in the reservoir and no particles on all other sites in the reservoir is
\begin{equation}
\hat{\mathcal{P}}_n(X) = \hat{a}^\dag_{x_1} \cdots \hat{a}^\dag_{x_n} |0\rangle_{R} {}_{R}\langle 0|\hat{a}_{x_n} \cdots \hat{a}_{x_1},
\end{equation}
where $|0\rangle_R$ is the direct product of vacuum states on all the reservoir sites. The final state after the measurement is thus
\begin{equation}
	| \Psi_{X}\rangle = \hat{\mathcal{P}}_n(X) | \Psi_{\text{BM}} \rangle.
	\label{eq:psi_m_general}
\end{equation}
Note that $|\Psi_{X} \rangle$ is not normalized. The norm
\begin{equation}\label{eq:probability}
P_n(X) = \langle \Psi_{X}|\Psi_{X}\rangle
\end{equation}
of the state is the probability to detect one particle on each of the reservoir sites $x_1, \dots, x_n$ and no particles on all other sites in the reservoir.

\section{\label{sec:measurement}Computation of expectation values}

We now show how expectation values of $|\Psi_{X} \rangle$ can be computed efficiently, despite the measurement. The derivation is not restricted to the Harper-Hofstadter model, but applies generally for noninteracting, fermionic Hamiltonians that preserve the number of particles. 

We will calculate the expectation value of $\hat{\phi}_a^\dag \hat{\phi}_b$ in the state $|\Psi_{X} \rangle$ after the measurement. Here, $\hat{\phi}_a^\dag$ and $\hat{\phi}_b^\dag$ denote general creation operators that act only within the system. In the position basis, they hence take the form
\begin{equation}
\hat{\phi}_i^\dag=\sum_{m\in S} \langle 0| \hat{a}_m \hat{\phi}_i^\dag |0\rangle \, \hat{a}^\dag_m, \quad i\in\{a,b\},
\end{equation}
where the sum runs only over the sites in the system. We also obtain an expression for the probability $\langle \Psi_{X}|\Psi_{X}\rangle$ to measure $n$ particles at specific positions in the reservoir.

Our starting point is the expression
\begin{equation}\label{Psi_X}
|\Psi_{X}\rangle = \hat{a}^\dag_{x_1} \cdots \hat{a}^\dag_{x_n} |0\rangle_{R} {}_{R}\langle 0 |\hat{a}_{x_n} \cdots \hat{a}_{x_1} \hat{\psi}_N^\dag \cdots \hat{\psi}_1^\dag|0\rangle.
\end{equation}
All terms in $\hat{\psi}_j^\dag$ that create particles on $R\setminus X$ will not contribute to $|\Psi_{X}\rangle$, and we can thus replace $\hat{\psi}_j^\dag$ by
\begin{equation}
\hat{\Tilde{\psi}}_j^\dag=\sum_{m \in S \cup X} \langle 0|\hat{a}_m \hat{\psi}_j^\dag|0\rangle\, \hat{a}^\dag_m.
\end{equation}
We can then rewrite $|\Psi_{X}\rangle$ further as
\begin{align}
|\Psi_{X}\rangle &= \hat{a}^\dag_{x_1} \cdots \hat{a}^\dag_{x_n} |0\rangle_{R} {}_{R}\langle 0 |\hat{a}_{x_n} \cdots \hat{a}_{x_1} \hat{\Tilde{\psi}}_N^\dag \cdots \hat{\Tilde{\psi}}_1^\dag |0\rangle\nonumber\\
&= \hat{a}^\dag_{x_1} \cdots \hat{a}^\dag_{x_n} |0\rangle_{X} {}_{X} \langle 0 |\hat{a}_{x_n} \cdots \hat{a}_{x_1} \hat{\Tilde{\psi}}_N^\dag \cdots \hat{\Tilde{\psi}}_1^\dag |0\rangle\nonumber\\
&= \hat{a}^\dag_{x_1} \cdots \hat{a}^\dag_{x_n} \hat{a}_{x_n} \cdots \hat{a}_{x_1} \hat{\Tilde{\psi}}_N^\dag \cdots \hat{\Tilde{\psi}}_1^\dag|0\rangle.
\end{align}
In the second line, we utilized that none of the operators create particles in $R\setminus X$, and in the third line we utilized that $\hat{a}_{x_n} \cdots \hat{a}_{x_1}$ ensures that all sites in $X$ are empty such that $|0\rangle_{X} {}_{X}\langle 0|$ acts as the identity operator. 

It is important to note that the $\{\hat{\Tilde{\psi}}_j^\dag \}$ operators do generally not constitute an orthonormal basis since we have removed the terms that create particles in $R\setminus X$. We therefore perform a Gram-Schmidt orthonormalization to arrive at the basis $\{ \hat{\eta}_i^\dag \}$ that satisfies $\{ \hat{\eta}_i , \hat{\eta}_j^\dag \} = \delta_{ij}$. This computation involves overlaps between single-particle states of dimension $|S|+n$, where $|S|$ is the number of lattice sites in the system, and the number of overlaps needed is determined by $N$ rather than by the number of lattice sites. For fixed $N$ and $n$, the computation time needed for the Gram-Schmidt orthonormalization is therefore a sum of a term proportional to $|S|$ and a term that is independent of the size of the lattice. In the resulting basis,
\begin{equation}
	\hat{\Tilde{\psi}}_j^\dag = \sum_{m=1}^{j} \langle 0|\hat{\eta}_m\hat{\Tilde{\psi}}_j^\dag|0\rangle \, \hat{\eta}_m^\dag.
\end{equation}
Note that the sum runs only up to $j$ by construction of the basis. Since the operators are fermionic, we have $(\hat{\eta}_m^\dag)^2=0$. Utilizing this, we get
\begin{multline}
| \Psi_{X} \rangle=
\hat{a}_{x_1}^\dag \cdots \hat{a}_{x_n}^\dag \hat{a}_{x_n} \cdots \hat{a}_{x_1} \hat{\eta}_N^\dag \cdots \hat{\eta}_1^\dag |0 \rangle\\
\times\prod_{j=1}^N \langle 0|\hat{\eta}_j\hat{\Tilde{\psi}}_j^\dag|0\rangle.
\end{multline}

Having expressed $| \Psi_{X} \rangle$ in terms of a product of fermionic operators, we compute 
\begin{multline}
\langle\Psi_{X}| \Psi_{X} \rangle=\\
\langle 0 |\hat{\eta}_1 \cdots \hat{\eta}_N \hat{a}_{x_1}^\dag \cdots \hat{a}_{x_n}^\dag \hat{a}_{x_n} \cdots \hat{a}_{x_1} \hat{\eta}_N^\dag \cdots \hat{\eta}_1^\dag |0 \rangle\\
\times\prod_{j=1}^N |\langle 0|\hat{\eta}_j\hat{\Tilde{\psi}}_j^\dag|0\rangle|^2
\end{multline}
and
\begin{multline}
\frac{\langle\Psi_{X}|\hat{\phi}_a^\dag \hat{\phi}_b | \Psi_{X} \rangle}{\langle\Psi_{X}| \Psi_{X} \rangle} =\\
\frac{\langle 0 |\hat{\eta}_1 \cdots \hat{\eta}_N \hat{a}_{x_1}^\dag \cdots \hat{a}_{x_n}^\dag \hat{\phi}_a^\dag \hat{\phi}_b \hat{a}_{x_n} \cdots \hat{a}_{x_1} \hat{\eta}_N^\dag \cdots \hat{\eta}_1^\dag |0 \rangle}{\langle 0 |\hat{\eta}_1 \cdots \hat{\eta}_N \hat{a}_{x_1}^\dag \cdots \hat{a}_{x_n}^\dag \hat{a}_{x_n} \cdots \hat{a}_{x_1} \hat{\eta}_N^\dag \cdots \hat{\eta}_1^\dag |0 \rangle}.
\end{multline}
In the second expression, we have utilized that $\hat{\phi}_a^\dag \hat{\phi}_b$ commutes with $\hat{a}_{x_n} \cdots \hat{a}_{x_1}$, since the former operator acts only on system sites, while the latter operator acts only on sites in the reservoir. Components of the $\hat{a}_{x_m}$ and $\hat{\phi}_i$ operators that annihilate particles outside the space spanned by the $\{\hat{\eta}_j^\dag\}$ operators will not contribute to the expectation values. We can therefore remove such components from the operators. This amounts to replacing $\hat{a}_{x_m}^\dag$ and $\hat{\phi}_i^\dag$ in the above expressions by
\begin{equation}
\hat{\tilde{a}}_{x_m}^\dag = \sum_{j=1}^{N} \langle 0| \hat{\eta}_j \hat{a}_{x_m}^\dag|0\rangle \hat{\eta}_j^\dag, \quad m \in \{1,2,\dots, n\},
\end{equation}
and
\begin{equation}
\hat{\tilde{\phi}}_i^\dag = \sum_{j=1}^{N} \langle 0|\hat{\eta}_j \hat{\phi}_i^\dag|0\rangle \hat{\eta}_j^\dag, \quad i\in\{a,b\}.
\end{equation}
These operators are not guaranteed to form an orthonormal basis. We thus perform another Gram-Schmidt orthonormalization to obtain a basis $\{ \hat{\xi}_j^\dag \}$, with $j\in\{1,2,\ldots,n\}$, satisfying $\{ \hat{\xi}_i, \hat{\xi}_j^\dag \} = \delta_{ij}$. The computation time for this operation is independent of $|S|$ and $|R|$, where $|R|$ is the number of sites in the reservoir, as one orthonomalizes $n+1$ single-particle states of dimension $N$. In the resulting basis,
\begin{equation}
\hat{\Tilde{a}}_{x_m}^\dag = \sum_{j=1}^{m} \langle 0|\hat{\xi}_j \hat{\Tilde{a}}_{x_m}^\dag|0\rangle \hat{\xi}_j^\dag, \quad m\in\{1,2,\ldots,n\},
\end{equation}
and
\begin{equation}
\hat{\Tilde{\phi}}_i^\dag = \sum_{j=1}^{n} \langle0|\hat{\xi}_j \hat{\Tilde{\phi}}_i^\dag |0\rangle \hat{\xi}_j^\dag + \langle0|\hat{\gamma}_i \hat{\Tilde{\phi}}_i^\dag |0\rangle \hat{\gamma}_i^\dag, \quad i\in\{a,b\}.
\end{equation}
Here, both of the $\hat{\gamma}_i^\dag$ operators anticommute with all the $\xi_j$ operators and $\{\hat{\gamma}_i,\hat{\gamma}_i^\dag\}=1$, but $\hat{\gamma}_a^\dag$ does not necessarily anticommute with $\hat{\gamma}_b$. Writing the operators in this basis, we obtain
\begin{equation}\label{eq:psiXpsiX}
\langle\Psi_{X} | \Psi_{X} \rangle = \prod_{m=1}^{n} |\langle 0|\hat{\xi}_m \hat{\Tilde{a}}_{x_m}^\dag|0\rangle|^2 \prod_{j=1}^N |\langle 0|\hat{\eta}_j\hat{\Tilde{\psi}}_j^\dag|0\rangle|^2
\end{equation}
and
\begin{widetext}
\begin{equation}\label{eq:expres}
\frac{\langle\Psi_{X} | \hat{\phi}_a^\dag \hat{\phi}_b | \Psi_{X} \rangle}{\langle\Psi_{X} | \Psi_{X} \rangle}
= \frac{\langle0|\hat{\gamma}_{a} \hat{\Tilde{\phi}}_a^\dag |0\rangle 
\langle0|\hat{\gamma}_{b} \hat{\Tilde{\phi}}_b^\dag |0\rangle^*
\langle 0 |\hat{\eta}_1 \cdots \hat{\eta}_N \hat{\xi}_{1}^\dag \cdots \hat{\xi}_{n}^\dag \hat{\gamma}_{a}^\dag \hat{\gamma}_{b}  \hat{\xi}_{n} \cdots \hat{\xi}_{1} \hat{\eta}_N^\dag \cdots \hat{\eta}_1^\dag |0 \rangle}{ \langle 0 |\hat{\eta}_1 \cdots \hat{\eta}_N \hat{\xi}_{1}^\dag \cdots \hat{\xi}_{n}^\dag \hat{\xi}_{n} \cdots \hat{\xi}_{1} \hat{\eta}_N^\dag \cdots \hat{\eta}_1^\dag |0 \rangle}
= \langle0|\hat{\gamma}_{a} \hat{\Tilde{\phi}}_a^\dag |0\rangle 
\langle0|\hat{\gamma}_{b} \hat{\Tilde{\phi}}_b^\dag |0\rangle^* \{\hat{\gamma}_b,\hat{\gamma}_a^\dag\}.
\end{equation}
\end{widetext}
The last equality follows because both $\hat{\gamma}_i$ operators act in the space spanned by the $\hat{\eta}_j^\dag$ operators minus the space spanned by the $\hat{\xi}_m^\dag$ operators, and this space is fully occupied by particles. One can thus replace $\hat{\gamma}_{a}^\dag \hat{\gamma}_{b}$ with $\hat{\gamma}_{a}^\dag \hat{\gamma}_{b}+\hat{\gamma}_{b} \hat{\gamma}_{a}^\dag$ inside the expectation value, as the second term contributes nothing anyway, and since the anticommutator is a number it can be taken outside. For the special case $a=b$, Eq.\ \eqref{eq:expres} simplifies to
\begin{equation}\label{eq:cor}
\frac{\langle\Psi_{X} | \hat{\phi}_a^\dag \hat{\phi}_a | \Psi_{X} \rangle}{\langle\Psi_{X} | \Psi_{X} \rangle}
= |\langle0|\hat{\gamma}_{a} \hat{\Tilde{\phi}}_a^\dag |0\rangle|^2.
\end{equation}

To evaluate the right hand side of Eq.\ \eqref{eq:psiXpsiX}, we need to multiply $(n+N)$ numbers already computed and take the norm squared, and to evaluate the right hand side of Eq.\ \eqref{eq:expres}, we need to multiply three overlaps between single-particle states of dimension $n+1$. This computation is hence also independent of $|S|$ and $|R|$. Altogether, for fixed $N$ and $n$, the time needed to compute the probability and the expectation values, starting from the single-particle operators $\{\hat{\psi}_j^\dag\}$ right before the measurement, thus scales only linearly with the number of sites in the system $|S|$. This enables us to do numerical computations for systems with a moderate number of particles and hundreds of sites for a given measurement outcome.

\section{Computation of averages over measurement outcomes}\label{sec:probability}

\subsection{Probability to measure $n$ particles in the reservoir}

In the previous section, we showed how to compute the probability $P_n(X)$ to detect particles on the sites $x_1,\ldots,x_n$ in the reservoir. With many sites in the reservoir, $P_n(X)$ is small, and one may also be interested in knowing the total probability $P_n$ to measure a total of $n$ particles in the reservoir. This probability is obtained by summing $P_n(X)$ over all possible choices of $\{x_1, \dots, x_n\}$, i.e.\
\begin{equation}
P_n=\sum_{X\in X_n} P_n(X),
\end{equation}
where $X_n$ is the set of all $\{x_1, \dots, x_n\}$ for which $x_1<x_2<\ldots<x_n$ and $x_j\in R$ for all $j$. The number of terms in this sum is $|R|!/[n!(|R|-n)!]$. This formula is, therefore, impractical to use for large reservoirs and $n$ not close to one. 

To reformulate the expression, we use \eqref{eq:probability} and \eqref{Psi_X} to obtain
\begin{multline}
P_n =\sum_{X\in X_n} \langle 0| \hat{\psi}_1 \cdots \hat{\psi}_N \hat{a}^\dag_{x_1} \cdots \hat{a}^\dag_{x_n} |0\rangle_{R}\\ \times {}_{R}\langle 0 | \hat{a}_{x_n} \cdots \hat{a}_{x_1} \hat{\psi}_N^\dag \cdots \hat{\psi}_1^\dag |0\rangle.
\end{multline}
We write
\begin{equation}\label{split}
\hat{\psi}_j =\hat{s}_j + \hat{r}_j =  \sum_{n_j=0}^1 \hat{r}_j^{n_j}\hat{s}_j^{1-n_j},
\end{equation}
where $\hat{s}_j$ is the part of $\hat{\psi}_j$ that acts on the system and $\hat{r}_j$ is the part of $\hat{\psi}_j$ that acts on the reservoir. Defining $\vec{n}=(n_1,\ldots,n_N)$ and $\vec{m}=(m_1,\ldots,m_N)$, this gives
\begin{multline}
P_n=\sum_{\vec{n}}\sum_{\vec{m}} 
\langle 0| \hat{r}_1^{n_1}\hat{s}_1^{1-n_1} \cdots \hat{r}_N^{n_N}\hat{s}_N^{1-n_N} \\
\times \sum_{X\in X_n} \hat{a}^\dag_{x_1} \cdots \hat{a}^\dag_{x_n} |0\rangle_{R}
{}_{R}\langle 0 | \hat{a}_{x_n} \cdots \hat{a}_{x_1} \\
\times (\hat{s}_N^{1-m_N})^\dag (\hat{r}_N^{m_N})^\dag \cdots (\hat{s}_1^{1-m_1})^\dag (\hat{r}_1^{m_1})^\dag |0\rangle\\
=\sum_{\vec{n}}\sum_{\vec{m}} \delta_{n,\sum_j n_j} \delta_{n,\sum_j m_j}
\langle 0| \hat{r}_1^{n_1}\hat{s}_1^{1-n_1} \cdots \hat{r}_N^{n_N}\hat{s}_N^{1-n_N} \\
\times (\hat{s}_N^{1-m_N})^\dag (\hat{r}_N^{m_N})^\dag \cdots (\hat{s}_1^{1-m_1})^\dag (\hat{r}_1^{m_1})^\dag |0\rangle,
\end{multline}
where we utilized that the second line acts as the identity for all states with $n$ particles in the reservoir, while all other states are removed. This can also be written as
\begin{equation}
P_n = \Tr(S_nR_n^T),
\end{equation}
where $S_n$ is the matrix with elements
\begin{equation}
S_{n,\vec{n},\vec{m}}=\langle 0|\hat{s}_1^{1-n_1} \cdots \hat{s}_N^{1-n_N}(\hat{s}^{\dag}_N)^{1-m_N} \cdots (\hat{s}^{\dag}_1)^{1-m_1}|0\rangle,
\end{equation}
and $R_n$ is the matrix with elements
\begin{multline}
R_{n,\vec{n},\vec{m}}=(-1)^{f(\vec{n})+f(\vec{m})}\\ 
\times \langle 0|\hat{r}_1^{n_1} \cdots \hat{r}_N^{n_N}(\hat{r}^{\dag}_N)^{m_N} \cdots (\hat{r}^{\dag}_1)^{m_1}|0\rangle
\end{multline}
with
\begin{equation}
f(\vec{n})=\sum_{j=1}^N jn_j.
\end{equation}
The matrices $S_n$ and $R_n$ both have $N!/(n!(N-n)!)$ rows and columns, namely all the possible choices of $\vec{n}$ under the constraint $\sum_{j}n_j=n$. Note that the dimension of the matrices $S_n$ and $R_n$ is independent of the size of the lattice and only depends on the number of particles in the system and reservoir.

Utilizing Wick's theorem, we get
\begin{equation}
S_{n,\vec{n},\vec{m}}=\det(S_O(\vec{n},\vec{m})).
\end{equation}
For a given $\vec{n}$ and $\vec{m}$, the overlap matrix $S_O(\vec{n},\vec{m})$ is the $(N-n)\times (N-n)$ matrix with elements $\langle 0| \hat{s}_i \hat{s}^\dag_j|0\rangle$, where only $i$ and $j$ values fulfilling $n_i=0$ and $m_j=0$ are included. Likewise,
\begin{equation}
R_{n,\vec{n},\vec{m}}=(-1)^{f(\vec{n})+f(\vec{m})}\det(R_O(\vec{n},\vec{m})),
\end{equation}
where $R_O(\vec{n},\vec{m})$ is the $n\times n$ matrix with elements $\langle 0| \hat{r}_i \hat{r}^\dag_j|0\rangle$, only including $i$ and $j$ values fulfilling $n_i=1$ and $m_j=1$.  

The number of lattice sites only enters through the computation of $\langle 0| \hat{s}_i \hat{s}^\dag_j|0\rangle$ and $\langle 0| \hat{r}_i \hat{r}^\dag_j|0\rangle$. Specifically, the time needed to compute $\langle 0| \hat{s}_i \hat{s}^\dag_j|0\rangle$ is proportional to $|S|$, and the time needed to compute $\langle 0| \hat{r}_i \hat{r}^\dag_j|0\rangle$ is proportional to $|R|$. For fixed $N$ and $n$, the time needed to compute $P_n$ is thus a sum of three terms of which one is proportional to $|S|$, another is proportional to $|R|$, and the third is independent of both $|S|$ and $|R|$.

\subsection{Average of expectation values}\label{sec:4B}

The expectation value of $\hat{\phi}_a^\dag \hat{\phi}_b$ averaged over all measurement outcomes with $n$ particles in the reservoir is 
\begin{multline}
\overline{\langle \hat{\phi}_a^\dag \hat{\phi}_b \rangle}_n = \frac{1}{P_n}\sum_{X\in X_n}\frac{\langle\Psi_X|\hat{\phi}_a^\dag \hat{\phi}_b|\Psi_X\rangle}{\langle\Psi_X|\Psi_X\rangle}P_n(X)\\
=\{\hat{\phi}_a^\dag,\hat{\phi}_b\}-\frac{1}{P_n}\sum_{X\in X_n} \langle\Psi_X|\hat{\phi}_b \hat{\phi}_a^\dag|\Psi_X\rangle.
\end{multline}
Following similar steps as above, we get
\begin{equation}
\sum_{X\in X_n}\langle\Psi_X|\hat{\phi}_b \hat{\phi}_a^\dag|\Psi_X\rangle = \Tr(S_{\phi,n} R_n^T),
\end{equation}
where $S_{\phi,n}$ is the matrix with elements
\begin{multline}
S_{\phi,n,\vec{n},\vec{m}}=\\
\langle 0|\hat{s}_1^{1-n_1} \cdots \hat{s}_N^{1-n_N}\hat{\phi}_b \hat{\phi}_a^\dag(\hat{s}^{\dag}_N)^{1-m_N} \cdots (\hat{s}^{\dag}_1)^{1-m_1}|0\rangle.
\end{multline}
These matrix elements can be expressed as a determinant using Wick's theorem as above. For fixed $N$ and $n$, the time needed to do the computation is again a term proportional to $|S|$ plus a term proportional to $|R|$ plus a term that is independent of both $|S|$ and $|R|$.

\subsection{Purity}

We next compute the purity of the system, when the reservoir is discarded without a prior measurement. The purity is defined as $\Tr(\rho_{\rm{S}}^2)$, where $\rho_{\rm{S}}$ is the reduced density operator of the system after tracing out the reservoir. Explicitly,
\begin{equation}\label{eq:rdo}
\rho_S=\sum_{R}\langle R| \hat{\psi}_N^\dag \cdots \hat{\psi}_1^\dag |0\rangle \langle 0| \hat{\psi}_1 \cdots \hat{\psi}_N |R\rangle,
\end{equation}
where the sum is over a complete set of states $|R\rangle$ of the reservoir. Hence,
\begin{multline}
\Tr(\rho_S^2)=\sum_{S}\sum_{R}\sum_{R'}\langle S|\langle R| \hat{\psi}_N^\dag \cdots \hat{\psi}_1^\dag |0\rangle \langle 0| \hat{\psi}_1 \cdots \hat{\psi}_N |R\rangle \\ \langle R'| \hat{\psi}_N^\dag \cdots \hat{\psi}_1^\dag |0\rangle \langle 0| \hat{\psi}_1 \cdots \hat{\psi}_N |R'\rangle|S\rangle,
\end{multline}
where $S$ is summed over a complete set of states $|S\rangle$ of the system. Utilizing \eqref{split}, we can write this as
\begin{multline}
\Tr(\rho_S^2)=\sum_{\vec{n}}\sum_{\vec{m}}\sum_{\vec{p}}\sum_{\vec{q}}
\sum_{S}\sum_{R}\sum_{R'}\\
\langle S|\langle R| (\hat{r}_N^\dag)^{n_N} (\hat{s}_N^\dag)^{1-n_N} \cdots (\hat{r}_1^\dag)^{n_1} (\hat{s}_1^\dag)^{1-n_1} |0\rangle\\ 
\times\langle 0| \hat{r}_1^{m_1} \hat{s}_1^{1-m_1} \cdots \hat{r}_N^{m_N} \hat{s}_N^{1-m_N}  |R\rangle \\ 
\times\langle R'|(\hat{r}_N^\dag)^{p_N} (\hat{s}_N^\dag)^{1-p_N} \cdots (\hat{r}_1^\dag)^{p_1} (\hat{s}_1^\dag)^{1-p_1} |0\rangle\\ 
\times\langle 0| \hat{r}_1^{q_1} \hat{s}_1^{1-q_1} \cdots \hat{r}_N^{q_N} \hat{s}_N^{1-q_N}  |R'\rangle|S\rangle\\
=\sum_{\vec{n}}\sum_{\vec{m}}\sum_{\vec{p}}\sum_{\vec{q}}\sum_{R}\sum_{R'}
(-1)^{f(\vec{n})+f(\vec{m})+f(\vec{p})+f(\vec{q})}\\
\langle 0| \hat{s}_1^{1-q_1} \cdots \hat{s}_N^{1-q_N} 
\hat{r}_1^{q_1} \cdots \hat{r}_N^{q_N}  |R'\rangle \\
\langle R| (\hat{r}_N^\dag)^{n_N} \cdots (\hat{r}_1^\dag)^{n_1} 
(\hat{s}_N^\dag)^{1-n_N} \cdots (\hat{s}_1^\dag)^{1-n_1} |0\rangle\\ 
\langle 0| \hat{s}_1^{1-m_1} \cdots \hat{s}_N^{1-m_N}  
\hat{r}_1^{m_1} \cdots \hat{r}_N^{m_N}|R\rangle \\ 
\langle R'|(\hat{r}_N^\dag)^{p_N} \cdots (\hat{r}_1^\dag)^{p_1} 
(\hat{s}_N^\dag)^{1-p_N} \cdots (\hat{s}_1^\dag)^{1-p_1}|0\rangle\\
=\sum_{\vec{n}}\sum_{\vec{m}}\sum_{\vec{p}}\sum_{\vec{q}}\\
\langle 0| \hat{s}_1^{1-q_1} \cdots \hat{s}_N^{1-q_N} (\hat{s}_N^\dag)^{1-n_N} \cdots (\hat{s}_1^\dag)^{1-n_1} |0\rangle\\
\times\langle 0|\hat{r}_1^{m_1} \cdots \hat{r}_N^{m_N} 
(\hat{r}_N^\dag)^{n_N} \cdots (\hat{r}_1^\dag)^{n_1} |0\rangle (-1)^{f(\vec{n})+f(\vec{m})}
\\ 
\times\langle 0| \hat{s}_1^{1-m_1} \cdots \hat{s}_N^{1-m_N}  
(\hat{s}_N^\dag)^{1-p_N} \cdots (\hat{s}_1^\dag)^{1-p_1}|0\rangle\\
\times\langle 0|\hat{r}_1^{q_1} \cdots \hat{r}_N^{q_N}  
(\hat{r}_N^\dag)^{p_N} \cdots (\hat{r}_1^\dag)^{p_1} |0\rangle (-1)^{f(\vec{p})+f(\vec{q})}\\
=\sum_{n=0}^N \Tr(S_nR_n^TS_nR_n^T).
\end{multline}
The elements of $S_n$ and $R_n$ can be expressed in terms of determinants as explained above. For fixed $N$ and $n$, the time needed to do the computation is a term proportional to $|S|$ plus a term proportional to $|R|$ plus a term that is independent of both.

\section{\label{sec:Results_and_discussion} Results and discussion}

\begin{figure*}
\includegraphics[width=\textwidth]{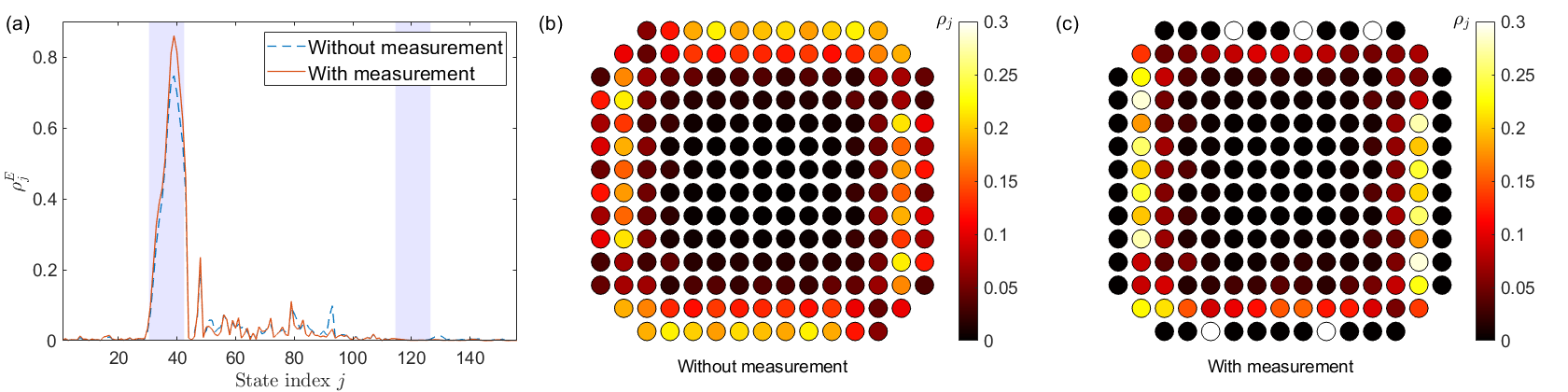}
\caption{Particle population of the energy eigenstates of the system and of the lattice sites for $T=\pi/(2J)$ and the measurement outcome illustrated in Fig.\ \ref{fig:scheme}(d). (a) The particle population of the energy eigenstates as a function of the state index. The blue background color indicates the clockwise (left) and counterclockwise (right) edge groups as given in Fig.\ \ref{fig:spectrum}. For this example, the total number of particles in the clockwise edge group is $R_{C} = 5.74$ without measurement and $R_{C} = 6.47$ with measurement. The corresponding numbers for the counterclockwise edge modes are $R_{CC} = 0.025$ and $R_{CC} = 0.024$, respectively. (b) The particle population of the sites when no measurement is performed. In this case, the state is the system is strongly mixed. (c) The particle population of the sites in the pure state right after the measurement. The white sites in the reservoir are occupied by one particle. As all other sites have a population below $0.3$, the color scale has been saturated to this value.}
\label{fig:example}
\end{figure*}

In the following numerical simulations, we use the system and reservoir illustrated in Fig.\ \ref{fig:scheme}. We take $\alpha = 1/4$, leading to the two groups of edge modes with opposite chirality shown in Fig.\ \ref{fig:spectrum}. We choose $\epsilon_R$ as in Eq.\ \eqref{er}, and we take $J_B$ to be zero, when the system and the reservoir are decoupled, and $J_B$ to be $J$ when they are coupled.  As explained above, we start out with $N=14$ particles in the reservoir to make their energies fit well inside the energy window of clockwise edge states. 

To quantify how well the proposed scheme succeeds to fill the edge states, we use Eq.\ \eqref{eq:cor} to numerically compute
\begin{equation}
\rho_j^E = \frac{\langle\Psi_{X} | \hat{\phi}_{j}^\dag \hat{\phi}_j | \Psi_{X} \rangle}
{\langle\Psi_{X} | \Psi_{X} \rangle},
\end{equation}
where $\hat{\phi}_j^{\dag}$ is the operator that creates a particle in the $j$th energy eigenstate of the system. The total population in the clockwise and counterclockwise edge states is then
\begin{equation}
R_C = \sum_{j\in C}\rho_j^E \quad \text{and} \quad R_{CC} = \sum_{j\in CC}\rho_j^E,
\end{equation} 
respectively. Another quantity of interest is the spatial particle density distribution, which is
\begin{equation}
\rho_j = \frac{\langle\Psi_{X} | \hat{a}_j^\dag \hat{a}_j | \Psi_{X} \rangle}
{\langle\Psi_{X} | \Psi_{X} \rangle},
\end{equation}
where $\hat{a}_j$ annihilates a particle on site $j$ in the system. This quantity is also computed using Eq.\ \eqref{eq:cor}.

We are here particularly interested in the effects of the measurement, and we shall therefore also compare to the case without the measurement. The results for the case without the measurement are obtained by making the replacement $|\Psi_X\rangle\to|\Psi_{\text{BM}}\rangle$ in the above equations. This gives the same result as first computing the reduced density operator for the system alone (see Eq.\ \eqref{eq:rdo}) and then computing the expectation value from the reduced density operator.

In Fig.\ \ref{fig:example}, we show an example for one of the most likely measurement outcomes among those with five particles in the reservoir. We have here chosen
\begin{equation}\label{eq:T}
T=\frac{\pi}{2J},
\end{equation}
which is the time needed for complete transfer when resonantly driving a two-level system with coupling strength $J$. We expect this to be a reasonable choice, as a single hop is sufficient to reach the system from the reservoir, but the actual dynamics is, of course, considerably more complex. Panel (a) shows that the majority of the particles are in the clockwise edge states after the measurement and that the population in the counterclockwise edge states is close to zero as desired. Note that it is more important to avoid particles in the counterclockwise edge states than to avoid particles in the bulk states, as particles in the counterclockwise and clockwise edge states both primarily occupy the edges of the lattice while particles in the bulk states primarily occupy the interior of the lattice. It is also seen that the number of particles in the clockwise edge states is higher after the measurement than without the measurement. Panels (b) and (c) show the populations of the lattice sites without and with the measurement, respectively. In both cases the highest populations are seen at the edges of the system and in the reservoir, and it is seen how the measurement projects the sites in the reservoir on being either empty or occupied by one particle. 

\begin{figure*}
\includegraphics[width=\linewidth]{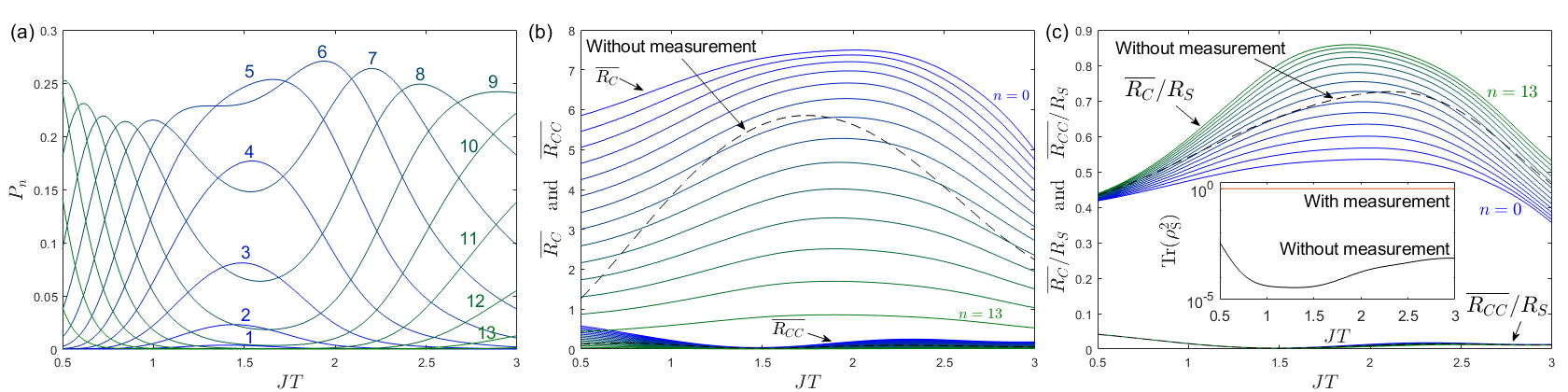}
\caption{(a) Probability $P_n$ to measure $n$ particles in the reservoir as a function of the time the system and reservoir are coupled. The individual curves are labeled by $n$. (b) The expectation value of the total number of particles in the clockwise (counterclockwise) edge states averaged over all measurement outcomes detecting a total of $n$ particles in the reservoir. For comparison, the dashed lines show the expectation value of the number of particles in the two sets of edge states when no measurement is performed. (c) The same data as in panel (b), except that we here divide by the expectation value of the total number of particles in the system. The inset shows the purity of the state of the system with and without the measurement. With measurement, the purity is unity, and without measurement, the purity is low because the reservoir is discarded without first disentangling it.}\label{fig:results}
\end{figure*}

To quantify how well the scheme works more generally, we plot averages of $R_C$ and $R_{CC}$ over all measurement outcomes with $n$ particles in the reservoir as a function of $T$ in Fig.\ \ref{fig:results}. The averages $\overline{R_C}$ and $\overline{R_{CC}}$ are computed by utilizing the results derived in Sec.\ \ref{sec:4B}. Panel (a) shows the probability to measure a total of $n$ particles in the reservoir, while panels (b) and (c) show the absolute and relative number of particles in the clockwise edge states, respectively. The relative populations are obtained by dividing $\overline{R_C}$ and $\overline{R_{CC}}$ by the total number of particles in the system
\begin{equation}
R_S = \sum_{j\in S}\rho^E_j.
\end{equation}
When $n$ particles are measured in the reservoir, we have $R_S=N-n$. For comparison, we also show $R_C$, $R_{CC}$, $R_C/R_S$, and $R_{CC}/R_S$ for the case without measurement.

We first discuss Fig.\ \ref{fig:results}(a). If the coupling has only been turned on for a short time, corresponding to a small value of $T$, one expects most of the particles to still be in the reservoir when the measurement is performed. This means higher probabilities for measuring relatively many particles in the reservoir, which results in relatively few particles in the system after the measurement. As $T$ increases, it becomes more likely that a larger number of particles have been transferred to the system before the measurement, and as $T$ reaches $\pi/(2J)$, the most probable outcome is to measure five particles in the reservoir, which leaves nine particles in the system. For even larger values of $T$, it again becomes more probable to measure a larger number of particles in the reservoir, as the particles that entered the system may also be transferred back to the reservoir again if the coupling stays on. We thus conclude that $T\approx \pi/(2J)$ is, indeed, a favorable choice to have a relatively large transfer. We also note that this value of $T$ is short compared to the coupling time $28/J$ considered in \cite{cold_atom_elevator}. The faster transfer is primarily due to the different choice of reservoir.

From the results in Fig.\ \ref{fig:results}(b) we observe that we generally end up with a larger number of particles in the clockwise edge states if we measure a smaller number of particles in the reservoir. This is expected as a smaller number of particles in the reservoir means a larger number of particles in the system. The additional particles in the system do, however, occupy the clockwise edge states to a lesser extent. As a result, the percentage of the particles in the system that are in the clockwise edge states after the measurement increases with increasing $n$ as seen in Fig.\ \ref{fig:results}(c). The increase is particularly large for $T\approx \pi/(2J)$. In other words, around this value of $T$, if many particles are measured in the reservoir, there are relatively few particles in the system, but most of those particles are in the clockwise edge states. If, instead, a small number of particles is measured in the reservoir, there is a larger number of particles in the system, but a larger fraction of them are not in the clockwise edge states. One can thus postselect on one or the other situation depending on whether it is most important to have many particles in the clockwise edge states or to have a large percentage of the particles in the clockwise edge states. Again, $T\approx \pi/(2J)$ or slightly larger is favorable as the maximum of $\overline{R_C}$ and of $\overline{R_C}/R_S$ appears in this range. Note also that $\overline{R_{CC}}$ and $\overline{R_{CC}}/R_S$ are generally small and are particularly small for $T\approx \pi/(2J)$. 

The results for the case without measurement are intermediate. This means that postselection on desired measurement outcomes can improve the results. As an example, for $T\approx \pi/(2J)$ it is relatively likely to measure three particles in the reservoir, which leads to a larger $\overline{R_{C}}$ than without the measurement. In addition, the projective measurement produces a pure state of the system instead of the highly mixed state obtained by discarding the reservoir without a prior measurement as seen in the inset of Fig.\ \ref{fig:results}(c).

\section{\label{sec:Conclusion} Conclusion}

We have investigated a scheme to populate topological, chiral edge states of a Chern insulator in the Harper-Hofstadter model. The initially empty system is coupled to a reservoir for a time $T$, and afterwards an instantaneous, projective measurement of the populations of the sites of the reservoir is performed. By doing the measurement, one obtains a pure state of the system. In addition, one can utilize postselection to obtain particularly desirable properties of the state of the system after the measurement. As an example, our computations suggest that a suitable strategy to obtain a particularly large number of particles in the clockwise edge states and a particularly small number of particles in the counterclockwise edge states is to choose $T\approx \pi/(2J)$ and postselect on detecting a small number of particles in the reservoir. If one rather wants to maximize the ratio between the number of particles in the clockwise edge states and the total number of particles in the system after the measurement, one should again choose $T\approx \pi/(2J)$, but instead postselect on measuring many particles in the reservoir. 

While the state right before the measurement can be computed based on single-particle physics, the measurement is more complicated to model, as it generally requires a full many-body description. We showed how one can nevertheless compute probabilities, expectation values, averages of expectation values, and purity efficiently for the state obtained after the measurement. Specifically, for a fixed number of particles, the time needed for the computations scales only linearly with the number of sites in the system. As long as the number of particles is moderate, computations can hence be carried out for setups with hundreds of sites despite the measurement step. This simplification is not limited to the Harper-Hofstadter model, but works generally for noninteracting, fermionic models with a fixed number of particles.

\begin{acknowledgments}
The authors thank Botao Wang for discussions and clarifications about their article \cite{cold_atom_elevator}. The work presented in this article is supported by Novo Nordisk Foundation grant NNF23OC0086670.
\end{acknowledgments}

\section*{Data Availability}
The source codes and parameters used to generate the figures are publicly available \cite{codes}.

\bibliography{refs}

\end{document}